\documentclass[aps,prd,onecolumn,nofootinbib,superscriptaddress,showpacs,amsmath,amssymb]{revtex4}
\usepackage{graphicx,epsf}
\usepackage[]{latexsym}
\newcommand{\ve}{\varepsilon}
\newcommand{\be}{\begin{eqnarray}}
\newcommand{\ee}{\end{eqnarray}}
\newcommand{\bea}{\begin{eqnarray}}
\newcommand{\eea}{\end{eqnarray}}

\def\comment#1{}

\newcommand{\lp}{\ell_{P}}
\newcommand{\mpl}{M_{P}}

\newcommand{\om}{\tilde{\omega}}

\usepackage{color}

\definecolor{darkred}{rgb}{.8,0,0}

\definecolor{darkblue}{rgb}{0,0,.7}

\definecolor{darkgreen}{rgb}{0,.7,0}

\begin{document}

%
%
\title{Generalized uncertainty principle and Asymptotic Safe gravity}

\author{Gaetano Lambiase}
\email{lambiase@sa.infn.it}
\affiliation{Dipartimento di Fisica "E.R. Caianiello", Universita' di Salerno, I-84084 Fisciano (Sa), Italy \&\\
INFN - Gruppo Collegato di Salerno, Italy}
\author{Fabio Scardigli\footnote{corresponding author}}
\email{fabio@phys.ntu.edu.tw}
\affiliation{Dipartimento di Matematica, Politecnico di Milano, Piazza Leonardo da Vinci 32, 20133 Milano, Italy}
\affiliation{Department of Applied Mathematics, University of Waterloo, Ontario N2L 3G1, Canada}
\begin{abstract}
\par\noindent

We present a procedure to link the deformation parameter $\beta$ of the generalized uncertainty principle (GUP) to the two free parameters $\om$ and $\gamma$ of the running Newtonian coupling constant of the Asymptotic Safe gravity (ASG) program. To this aim, we compute the Hawking temperature of a black hole in two different ways. The first way involves the use of the GUP in place of the Heisenberg uncertainty relations, and therefore we get a deformed Hawking temperature containing the parameter $\beta$. The second way involves the deformation of the Schwarzschild metric due to the Newtonian coupling constant running according to the AS gravity prescription. The comparison of the two techniques yields a relation between $\beta$ and $\om$, $\gamma$. As a particular case, we discuss also the so called $\xi$-model. The relations between $\beta$ and $\om$, $\xi$ allow us to transfer upper bounds from one parameter to the others.
\end{abstract}
%

%
\maketitle
\section{Introduction}

It is well-known that the formulation of a consistent quantum theory of gravity is still an open task in modern theoretical physics. Many of the approaches to the problem existing today in the literature (for a partial list see e.g. \cite{QG1,QG2,QG3,QG4,QG6,QG8,QG9} and references therein), share one particular property. Namely, the basic parameters that enter into the action defining the model at hand, such as Newton's constant, electromagnetic coupling, the cosmological constant etc, become scale dependent quantities. This does not come as a surprise of course, since scale dependence at the level of the effective action is a generic feature of ordinary quantum field theory. In theories of gravity, the scale dependence is expected to modify the horizon, the thermodynamics as well as the quasinormal modes spectra of classical black hole backgrounds
\cite{Koch:2016uso,
Rincon:2017goj,
Contreras:2017eza,Rincon:2018sgd,Contreras:2018dhs,Rincon:2018lyd,
Rincon:2018dsq,
Rincon:2019cix,
Contreras:2018gpl}. Also, the Sagnac effect \cite{Rincon:2019zxk}, the evolution of trajectories of photons \cite{Fathi:2019jid}, some cosmological solutions \cite{Canales:2018tbn}, and transverse wormhole solutions \cite{Contreras:2018swc} have been investigated.
In the context of theories beyond classical GR, apart from the aforementioned approach based on scale-dependent gravity, there is yet another method, which is usually called "improved" asymptotically safe (AS) gravity \cite{Bonanno:2000ep,Bonanno:2001xi,Reuter:2003ca}. For RG-improved cosmologies and inflationary models from asymptotic safety see e.g. \cite{Platania:2020lqb,Bonanno:2002zb,Bonanno:2001hi,Liu:2018hno,Hindmarsh:2012rc}, and for recent progress \cite{Platania:2019kyx,Moti:2018uvl,Koch:2013owa,Bonanno:2016dyv,chinosII,Li:2013kkb}.
%
%
In that scenario, the main idea is to integrate the beta function for the gravitational coupling in order to compute Newton's constant as a function of some energy scale $k$, $G(k)$.  After that, Newton's constant is inserted into the classical black hole solution and the improved lapse function is obtained. It is essential to notice that the gravitational coupling depends on some arbitrary renormalization scale $k$. Finally, a link between the energy scale $k$ and the radial coordinate $r$ is established. Only after this final step, the complete form of an improved black hole solution is obtained. Those extended solutions, inspired by the asymptotic safety program, are expected to modify the classical black hole solutions by incorporating quantum features. Similar black hole solutions, free of the central singularity, can be found, e.g, in Refs.~\cite{bardeen,hayward,frolov}, although they are not directly connected with the ASG program.\\
\\
Another important stream of ideas that try to describe the interplay between quantum effects and gravity is the research line known under the name of Generalized Uncertainty Principle (GUP). Although some seminal work on possible generalizations of the fundamental Heisenberg Uncertainty Principle (HUP)~\cite{Heisenberg} can be traced back to more than seventy years ago \cite{GUPearly}, these concepts have been intensely revived and precised in the last thirty years \cite{VenezGrossMende,MM,kempf,FS,Adler2,CLS,SC}. In a nutshell, several re-analysis of the measurement process, in particular of the famous Heisenberg microscope argument, have shown, with plenty of details, that, when the action of gravity is taken into account, the fundamental uncertainty relation should be modified as
\be
\mbox{\hspace{-5mm}}\Delta x\, \Delta p &\geq&
\frac{\hbar}{2}\left(1  +  \beta\,\frac{4\,\lp^2}{\hbar^2}\,\Delta p\,^2\right) 
=
\frac{\hbar}{2}\left[ 1 + \beta \left(\frac{\Delta p}{\mpl} \right)^2\right]
\label{gup}
\ee
Here $\lp$ is the Planck length, $\mpl$ the Planck mass, and we work in units where $2G_0 \mpl=\lp$, $\hbar=2\mpl\lp$, $c=k_B=1$.
As said, studies in this framework focus on understanding the effects of gravity on the formulation of uncertainty relations. Therefore, it turns out natural that the most relevant modifications to the HUP have been proposed in contexts such as string theory, non commutative geometry, and studies of black hole physics~\cite{VenezGrossMende,MM,FS,Adler2,kempf,CLS,SC}. The dimensionless parameter $\beta$ in (\ref{gup})
is not fixed \textit{a priori} by the theory, however in some models of string theory it is generally assumed that $\beta \sim {\cal O}(1)$ \cite{VenezGrossMende}.

The aim of this paper is to present and discuss a link between the free parameters of ASG, i.e. the renormalization scale, and the deforming parameter $\beta$ of the generalized uncertainty principle (GUP) .
%
%
Here we shall use the GUP to compute the Hawking temperature of a given black hole, obtaining it as a function of the deformation parameter $\beta$. The same deformed temperature can however be computed as well via the ASG deformation of Schwarzschild metric. This will yield a relation between $\beta$ and the ASG free parameters.

The paper is organized as follows. In the next Section we shortly recall the main results of the renormalization group approach leading to the ASG-improved classical Newton potential, as well to the ASG-modification of the Schwarzschild metric. In Section III we compute the GUP-deformed Hawking temperature. In Section IV we compute the ASG-deformed Hawking temperature. In Section V we compare such two deformations and relate the respective deformation parameters. In Section VI we give a precise evaluation of the ASG parameters. Section VII is devoted to discuss a particular example. Conclusion are shortly presented in the last Section.

\section{Running Newtonian coupling and Black hole metrics}
\label{Classical}
\noindent

As we briefly summarized in the Introduction, the main steps towards the construction of a renormalization group improved Schwarzschild metric are essentially three (we make particular reference to Ref.~\cite{Bonanno:2000ep}): first, we integrate the beta function for the gravitational coupling to compute Newton's constant as a function of some energy scale $k$, namely $G(k)$. It is essential to notice that the gravitational coupling depends on some arbitrary renormalization scale $k$. After that, a link between the energy scale $k$ and the radial coordinate $r$ must be established (the so called "identification of the infrared cutoff", namely $k=k(r)$). Finally, the $G(r)$ Newton's constant is inserted into the classical black hole solution and we obtain the improved lapse function of the metric. It is only after this final step that the complete solution of a "renormalization improved" black hole becomes concretely usable for explicit calculations. The above steps are detailed in Appendix A.

So, the basic idea of the AS gravity approach in order to obtain the renormalization improved, classical Newtonian or general relativistic, solutions is to replace everywhere the Newton constant $G_0$ with the running constant $G(r)$, whose explicit form is given by \cite{Bonanno:2000ep}
\be
\label{Geffr}
G(r) = \frac{G_0 r^3}{r^3 + \om G_0 \hbar \left(r + \gamma G_0 M \right)} \,,
\ee
where, in accordance with our conventions, $c=k_B=1$ and we retained $\hbar$. Here $\om$ and $\gamma$ are dimensionless numerical parameters, whose concrete value will be discussed later.

The line element for the spherically symmetric, Lorentzian metric preserves the usual form, that is
\be
\mathrm{d}s^2 = f(r)\mathrm{d} t^2 - f(r)^{-1} \mathrm{d} r^2 - r^2 \: \mathrm{d}\Omega^2,
\label{metric}
\ee
where $r$ is the radial coordinate, and $\mathrm{d} \Omega^2 = \mathrm{d \theta^2 + sin^2 \theta \: d \phi^2}$ is the line element of the unit two-sphere. But now, according to the above prescriptions, the lapse function $f(r)$ of our ASG improved Schwarzschild geometry reads
\be
\label{ASGmetric}
f(r) = 1 - \frac{2 M G(r)}{r}=1-\frac{2 G_0 M r^2}{r^3 +  \om G_0 \hbar (r + \gamma G_0 M)}\,,
\ee
with $G(r)$ given by (\ref{Geffr}) and $M$ the mass of the black hole. Of course, we suppose $\om \neq 0$, otherwise we would go back to the standard Schwarzschild metric.
Two very important limiting cases should be considered. \\
The first corresponds to the low energy scales ($r \rightarrow \infty$, or $k \to 0$), which implies
\be
f(r \rightarrow \infty) \simeq 1 - \frac{2 G_0 M}{r} ,
\label{infty}
\ee
so we recover the standard Schwarzschild metric at large distances, and this behavior is independent from the values of $\om$ and $\gamma$. \\
The second corresponds to the high energy scales ($r \rightarrow 0$, or $k \to \infty$). Here we have to distinguish two subcases:\\
if $\gamma \ne 0$, then
\be
f(r \rightarrow 0) \simeq 1 - \frac{2 r^2}{\om \gamma G_0 \hbar}\,,
\label{zero}
\ee
and thus the lapse function corresponds to a deSitter ($\om\gamma > 0$) or an Anti-deSitter ($\om\gamma <0$) core of our metric, depending on the sign of $\om\gamma$;\\
if $\gamma = 0$, then
\be
f(r \rightarrow 0) \simeq 1 - \frac{2 M r}{\om \hbar}\,,
\ee
so in this case we have a so-called conic singularity at the origin.
Clearly, the $\hbar$ at denominator is the hallmark of the quantum character of the correction that the ASG approach gives to the core of the standard Schwarzschild metric. Note that in both cases the central singularity is wiped out.

\section{GUP-deformed Hawking temperature}

The argument based on Heisenberg microscope~\cite{Heisenberg} suggests that
the size $\delta x$ of the smallest detail of an object (theoretically detectable
with a beam of photons of energy $E$) is roughly given by
\be
\delta x
\simeq
\frac{\hbar}{2\, E}
\label{HS}
\ee
since ever larger energies are required to explore ever smaller details.
On the other hand, the uncertainty relation~\eqref{gup} implies that the GUP version of Eq. \eqref{HS} is
\be
\delta x
\simeq
\frac{\hbar}{2\, E}
+ 2\,\beta\,\lp^2\, \frac{E}{\hbar}
\label{He}
\ee
which relates the (average) wavelength of a photon to its energy $E$.
The thermal GUP corrections to the Hawking spectrum are traced in many references \cite{FS9506,ACSantiago,CDM03,Susskind,nouicer,Glimpses}.
Equation (\ref{He}) allow to derive a relation between the mass and the temperature of a Schwarzschild black hole.
To this aim, consider an ensemble of unpolarized photons of Hawking radiation just outside the event horizon of a Schwarzschild black hole.
The position uncertainty of such photons can be estimated to be $\delta x  \simeq  2\mu R_S$, where $R_S=2G_0 M$ is the Schwarzschild radius of the black hole. We fixed $\mu = \pi$ to recover the standard semiclassical
Hawking temperature when $\beta \to 0$ (see below). The equipartition principle imposes that the average energy
$E$ of unpolarized photons of the Hawking radiation is related to the temperature $T$ as
$E \ \simeq \ T$, so that Eq. (\ref{He}) yields
\be
M \ = \ \frac{\hbar }{8\pi G_0  T} \ + \ \beta\, \frac{ T}{2\pi }\,.
\ee
Inverting this relation, and expanding in $\beta$, we obtain
\be
T(M) = \frac{\hbar}{8\pi G_0 M} \left(1 + \frac{\beta\,\mpl^2}{4\pi^2\,M^2} + \dots \right)~.
\label{Tg}
\ee
In deriving (\ref{Tg}) we have assumed the thermal character of the correction induced by the GUP.
There are, however, different approaches in which the corrections do not respect the exact thermality of the spectrum, and thus need not be reducible to a simple shift of the temperature (an example is the corpuscular
model of a black hole of Refs.~\cite{dvali}). Further aspects of the GUP are discussed (among the many papers) in Refs.~\cite{GUP}.

\section{Hawking temperature from the ASG-deformed Schwarzschild metric}

Of course, the Bekenstein-Hawking temperature for the general class of black holes with metrics of type \eqref{metric} can be computed with a well known general procedure \cite{Hawking} for any arbitrary function $f(r)$. By assuming that $f$ has a simple zero at some $r_h$, namely $f(r_h)=0$ and $f'(r_h)\neq 0$, and that it increases monotonically from zero to $f(+\infty)=1$ for $r>r_h$, then the Bekenstein-Hawking temperature on the horizon $r=r_h$ is
\be
T_{BH}=\frac{\hbar}{4 \pi} \, f'(r_h)\,.
\label{TBH}
\ee
Here the above calculation can be in principle performed exactly, using for $f(r)$ the expression given in Eq.\eqref{ASGmetric}, although a cubic equation would be involved (see Ref.\cite{Bonanno:2000ep} for this procedure). However, to our scope, since we have in mind to match the ASG-deformed temperature with the GUP-deformed temperature in order to extract the relation between $\beta$ and $(\om,\gamma)$, it results much more doable and clear to write $f(r)$ of Eq.\eqref{ASGmetric} in the form of a small perturbation over the Schwarzschild metric (we follow  \cite{SLV,SC2,LambSME,SLVConf,tomoyo})
\be
f(r) = 1 - \frac{2G_0 M}{r} + \frac{2 \,G_0^2 \,\om \,M \hbar \,(r + \gamma G_0 M)}{r^4} + {\cal O}(\frac{1}{r^5})
= 1 - \frac{2G_0 M}{r} + \ve(r)
\label{ve}
\ee
with
\be
\ve(r) \simeq \frac{2 \,G_0^2 \,\om \,M \hbar \,(r + \gamma G_0 M)}{r^4}\,,
\label{ver}
\ee
and this can be done since $|\ve(r)|\ll2G_0 M/r$ for any $r>2G_0M$.
For any metric of the form \eqref{ve} the horizon equation $f(r)=0$ reads
\be
r - 2G_0M + r\ve(r) = 0\,,
\ee
which can be solved to the first order in $\ve(r)$ by
\be
r_h = a - \frac{a\ve(a)}{1+\ve(a)+a\ve'(a)} = a\left[1-\ve(a) + {\cal O}(\ve^2)\right]\,,
\label{rh}
\ee
where $a=2G_0M$. Using now Eqs.\eqref{TBH}, \eqref{ve}, and \eqref{rh}, we can expand in $\ve$ and arrive to the deformed Hawking temperature
\be
T_{BH}=\frac{\hbar}{4 \pi} \, f'(r_h) =
\frac{\hbar}{4\pi a}\left\{1 + [2\ve(a) + a\ve'(a)] + \ve(a)[\ve(a)-2a\ve'(a)-a^2\ve''(a)] + {\cal O}(\ve^3)\right\}
\label{teps}
\ee
where for sake of completeness we report also the second order in $\ve$. The symbol ${}^\prime$ stands for the derivative with respect to $r$ (${}^\prime \equiv \partial_r$).
%
%

\section{Relation between $\beta$ and $\om, \gamma$}

By comparing the first orders of the expansions of the GUP-deformed Hawking temperature given in Eq.(\ref{Tg}), and of the ASG-Schwarschild temperature (\ref{teps}), one obtains
\be
\label{beta1}
\beta  =   \frac{4\pi^2 M^2}{\mpl^2}\, \left[2\ve(a) + a \ve'(a)\right]\,,
\ee
where $a=2G_0 M$. Finally, using the expression (\ref{ver}) for $\ve(r)$ we arrive at
\be
\label{beta}
\beta \ = \ -\,4\, \pi^2 \, \om \, (1+\gamma)\,.
\ee
This is the general relation between the ASG parameters $\om$, $\gamma$ and the deformation parameter $\beta$ of the GUP.

\section{Estimate of parameters $\om$, $\gamma$, $\beta$}

As we said in Sec.II, the ASG-improved Newtonian potential can be obtained from the standard Newton formula
\be
V(r) = - \frac{G_0 M m}{r}
\ee
by simply replacing the experimentally observed Newton constant $G_0$ with the running coupling $G(r)$ given in Eq.\eqref{Geffr}. Thus we get
\be
V^{ASG}(r) = -\frac{G(r)Mm}{r} = -\frac{G_0 \,M\, m\, r^2}{r^3 + \om \,G_0 \,\hbar \,(r + \gamma G_0 M)}\,,
\ee
which can be expanded for large $r$ as
\be
\label{Newrun}
V^{ASG}(r) = - \frac{G_0 M m}{r}\left[1 - \frac{\om G_0 \hbar}{r^2} - \frac{\gamma \om G_0^2 \hbar M}{r^3} + {\cal O}\left(\frac{G_0^2 \hbar^2}{r^4}\right)\right]\,.
\ee
We should notice that the corrections to the standard Newtonian potential predicted by the ASG approach are all of quantum nature. This is clearly suggested by the presence of $\hbar$ in each term of correction. In fact, there are no correction terms of classical origin, coming from some kind of post-Newtonian approximation.
On the other hand, corrections of quantum origin to the classical Newtonian potential have been elaborated by several researchers \cite{Duff,donoghue} in the last three decades or so. In particular, it was pointed out by Donoghue \cite{donoghue,donoASP} that the standard perturbative quantization of Einstein gravity leads to a well defined, finite prediction for the leading large distance quantum correction to Newtonian potential. The numerical coefficients of the quantum expansion have undergone a certain evolution over the years~\cite{kiril}, but the result today accepted by the community~\cite{dono2} reads
\be
\label{Newquan}
  V^{Quantum}(r) =
	- \frac{G_0 M m}{r}\left[1 \ + \ \frac{41}{10\pi}\frac{G_0 \hbar}{r^2} \ + \ \dots\right]\,.
\ee
This is an expansion at first order in $\hbar$ (up to classical terms due to non-linear nature of General Relativity, which can be reabsorbed by a change of coordinates), where the first correction term represents a genuine quantum correction proportional to $\hbar$. 

The comparison of the two expansions \eqref{Newrun} and \eqref{Newquan} fixes the parameter $\om$, which results to be
\be
\om = - \frac{41}{10\pi}\,.
\label{om}
\ee
The ASG parameter $\gamma$ is not fixed by these considerations. In this regard, we can follow two different paths:
\begin{description}
\item[$\textbf{i)}$] We can refer to the considerations developed in Ref.\cite{Bonanno:2000ep}. Those classical general relativistic arguments fix $\gamma = 9/2$. Using this value of $\gamma$ and the Eqs.\eqref{beta},\eqref{om}, we can compute the $\beta$ parameter of the GUP, which results to be
\be
\beta = \frac{451 \, \pi}{5}\,.
\ee
As we see $\beta$ results to be of order $10^2$ as suggested by some models of string theory \cite{VenezGrossMende}.
\item[$\textbf{ii)}$] We can first compute $\beta$, by following the procedure described in Ref.\cite{SLV}, namely comparing the GUP Hawking temperature \eqref{Tg} with the Hawking temperature of a Schwarzschild metric corrected through the Donoghue potential \eqref{Newquan}. This yields $\beta=82\pi/5$. Then, using \eqref{beta},\eqref{om}, we get
\be
\gamma = 0\,.
\ee
\end{description}

Given the indeterminacy of the arguments used in \cite{Bonanno:2000ep} to fix $\gamma$, we would incline for the second possibility.
Finally, the study of the metric \eqref{ASGmetric}, and of the related (somehow surprising) black hole features when $\om<0$, will be carried out in Ref.~\cite{SLnew}.



\section{The $\xi$-model}

In this Section we investigate further an example of modified, or "improved", Schwarzschild solution, described by the lapse function
\be
\label{xidef}
f(r) = 1 - \frac{2G_0 Mr}{r^2 + \xi}
\ee
Mathematically, the above lapse function is just a particular case of the previous \eqref{ASGmetric} with $\gamma=0$ and the identification $\xi = \om G_0 \hbar$.
Metrics as the above have been studied for example in Refs. \cite{chinosII,Li:2013kkb} as an approximate expression of the lapse
\be
f(r) = 1 - \Lambda \eta - \frac{\Lambda r^2}{3} - \frac{2G_0 Mr}{r^2 + \xi}\,,
\label{xi2}
\ee
where the parameters $\eta$ and $\xi$ have the dimensions of a length squared and they are supposed, at least initially, to be positive.
The authors obtained the lapse \eqref{xi2} as the static, spherically symmetric solution (in the infrared limit) of an ASG quantum-corrected black hole in an Anti-de Sitter (AdS) spacetime, with a deficit angle $\Lambda \eta$. However, the effect of the latter is negligibly small, as referred to cosmological observations and, for an astrophysical black hole, the corrections yielded by the cosmological constant are extremely small. Therefore, neglecting $\Lambda$, the authors arrive to the expression \eqref{xidef}. At a far distance the lapse \eqref{xidef} correctly reproduces the Schwarzschild solution (recovered for $\xi=0$), and deflection from it is of the order of $\xi$. At the origin the metric is finite, however it is not regular, since it contains a conic singularity. According to Ref.\cite{frolov}, the metric has a curvature invariant $R$ of the form
\be
R = \frac{2G_0M}{\xi^{3/2}}g(r/\sqrt{\xi})
\ee
(where $g$ is a finite function), which means that for any fixed $r$ the corresponding $R$ can be made arbitrarily large by simply increasing the mass parameter $M$. Therefore the metric \eqref{xidef} for a black hole does not satisfy the so-called limiting curvature condition. Of course, the same can be said for the lapse \eqref{ASGmetric} in the
case $\gamma=0$.

From (\ref{xidef}) we can expand for large $r$ (the regime $r^2 \gg \xi$), and get
\be
f(r) = 1 - \frac{2G_0M}{r} + \frac{2G_0M\xi}{r^3} + {\cal O}\left(\frac{1}{r^5}\right)\,,
\label{xiexp}
\ee
so we can identify the deformation to the Schwarzschild metric, i.e.
\be
\ve(r) \simeq \frac{2G_0M\xi}{r^3}\,.
\ee
Thus Eq. (\ref{beta1}) gives
\be
\label{betaxi}
\beta=-\frac{\pi^2 \xi}{G_0^2 \mpl^2} = -\frac{4 \pi^2 \xi}{\lp^2}\,.
\ee
As arises from Eq. (\ref{betaxi}), the GUP parameter $\beta$ is independent on the gravitational mass source, and it is only related to the fundamental Planck mass-scale. In case we assume $\xi > 0$ \textit{a priori}, then $\beta$ would result negative. This would not sound particularly weird, if we remember that a negative $\beta$ appears in contexts as uncertainty principle on a lattice \cite{JKS}, or white dwarf stability \cite{ong} or  decoherence limit of quantum systems (Tsallis thermostatistics) \cite{jizba2022}.

However, if we coherently follow the logic of the previous section, then from \eqref{xiexp} we can extract the expansion of the effective Newtonian potential $V_N=[f(r)-1]/2$, and by comparing it with the Donoghue quantum-corrected Newtonian potential \eqref{Newquan}, we get
\be
\xi = -\frac{41}{10\pi}\,G_0\hbar = -\frac{82}{5\pi}\,G_0^2\mpl^2
\ee
which inserted in \eqref{betaxi} yields
\be
\beta = \frac{82\pi}{5}\,,
\ee
consistently with the case $ii)$ of the previous section ($\gamma=0$).

Two comments are now in order.
\begin{itemize}
\item
Requiring the matching with the Donoghue quantum-corrected Newtonian potential at large $r$, the parameter $\xi$ results actually to be \textit{negative}, as it resulted also the parameter $\om$ (see the previous section), consistently with the identification $\xi = \om G_0 \hbar$. This fact has heavy consequences on the structure of the metric \eqref{xidef} for singularities, horizons, etc. This will form the subject of a forthcoming paper.
\item
In the same vein as in Ref.\cite{LambSME}, we can use the relations \eqref{beta} (with $\gamma=0$) and \eqref{betaxi}, connected by $\xi = \om G_0 \hbar$, in order to transfer experimental upper bounds from the GUP parameter $\beta$ to the ASG parameters $\om$, $\xi$.
\end{itemize}



From \eqref{beta} it is clear that between $\beta$ and $\om$ there is a mere factor $\sim 40$, therefore upper bounds on $\beta$ are essentially reflected directly on $\om$. Instead, using \eqref{betaxi} in the form $|\xi| = (\lp/2\pi)^2 |\beta|$, we obtain for the upper bounds on $\xi$ the results reported in Table \ref{Tab1}.
Of course, if the GUP parameter is of the order of unity, $\beta\sim {\cal O}(1)$, then $\xi$ is of the order of the Planck Length squared, that is
 \[
\xi\sim \lp^2\,.
 \]

\begin{center}
\begin{table}[ht]
\caption{Upper bounds on $\xi$ derived from gravitational and non-gravitational experiments.}
\begin{tabular}{c|c|c}
  \hline\hline
  $\beta <$ & Physical framework   & $\sqrt{\xi} \,\, cm <$ \\ \hline
%
%
 $10^{6}$  &  Sapphire mechanical resonator \cite{Bushev} & $10^{-31}$ \\ \hline
 $10^{12}$ & Micro and nano mechanical   & $10^{-28}$ \\
             & oscillators (masses $\sim \mpl$) \cite{Bonaldi} & \\ \hline
 $10^{20}$ & Lamb shift   \cite{vagenas} &  $10^{-24}$ \\ \hline
 $10^{21}$ & Scanning tunneling microscope \cite{vagenas} & $10^{-23}$ \\ 
           & Violation of equivalence principle \cite{ghosh} &  \\ \hline
 $10^{33}$   & Gravitational bar detectors  \cite{martin} &  $10^{-18}$ \\ \hline
 $10^{34}$  & Electroweak measurement \cite{SC2} & $10^{-17}$ \\ 
            & Charmonium levels \cite{vagenas}   & \\
            & Energy difference in Hydrogen levels $1S-2S$ \cite{Tkachuk} & \\ \hline
 $10^{39} $ & ${}^{87}$Rb cold-atom-recoil experiment  \cite{gao} & $10^{-14}$ \\ \hline
 $10^{46}$ & Landau levels  \cite{vagenas} & $10^{-11}$ \\ \hline\hline
 \end{tabular}
\label{Tab1}
\end{table}
\end{center}

As a general comment, we can say that it is important, from an experimental point of view, to have revealed a direct connection between the parameters $\beta$, $\om$, $\xi$, namely Eqs.\eqref{beta},\eqref{betaxi}. This allows us to transfer experimental upper bounds from one parameter to the other, thus enriching, at least in principle, the phenomenology available to researchers in the field.

\section{Conclusions}

In this paper we have derived an exact relation between the GUP parameter $\beta$ and the parameters $\om, \gamma$ characterizing the Asymptotic Safe gravity. The shift of the Hawking black hole temperature, for which the GUP is relevant, is derived by means of pure quantum mechanical principles, and no specific representation of the canonical commutation relation is postulated. On the other hand, the same temperature is derived geometrically for an ASG-improved version of the Schwarzschild metric, allowing to link the deformed uncertainty relation parameter $\beta$ with the renormalization scale, here measured by the parameters $\om, \gamma$ (or $\xi$, in a specific example).

As a byproduct, from the well known and up-to-date upper bounds on GUP parameter $\beta$ present in literature, we obtained an upper bound on the renormalization scale parameters $\om, \xi$ (derived in the framework of ASG running coupling Newtonian constant).
%
Indeed, a common point in the wide discussions on various models yielding the GUPs is related to test the size of these modifications of the uncertainty relations. These aspects appear particularly interesting in the perspective of a laboratory-scale imitation of the black hole horizon, with the subsequent possible emission of an analogue Hawking radiation \cite{ron}.

Aside from the above results, through a comparison with the quantum corrected Newtonian potential, we have also determined the exact numerical value of the ASG parameters $\om, \xi$, which both result to be negative. The deep consequences that this fact has on the related black hole metrics will be explored in Ref.~\cite{SLnew}.

\appendix
\section{Renormalization group-improved Schwarzschild solutions}

%
%

In this Appendix we shall review the improved Schwarzschild solution obtained in \cite{Bonanno:2000ep}.
The average Einstein-Hilbert action, able to avoid ghosts, is given by ($\hbar=1$)
\begin{align}
\Gamma_k[\mathfrak{g}] &= \frac{1}{16 \pi G(k)} \int \mathrm{d}^4 x \sqrt{\mathfrak{g}} \Bigl( - R(\mathfrak{g}) + 2 \bar{\lambda}(k)\Bigl).
\end{align}
Here $\mathfrak{g}$ is the metric, while $G(k)$ and $\bar{\lambda}(k)$ are the running Newton's constant and the cosmological constant, respectively.
The evolution of the scale-dependent couplings is governed by the Wetterich equation
\begin{align}\label{watt}
\partial_{t}\Gamma_{k} &= \frac{1}{2} \text{Tr}\left(\frac{\partial_{t}\mathcal{R}_{k}}{\Gamma^{(2)}_{k}[\phi] + \mathcal{R}_{k}}\right),
\end{align}
where $t=\ln(k)$, $\Gamma_{k}^{(2)}$ is the Hessian of $\Gamma_k$ with respect to $g_{\mu \nu}$, and $\mathcal{R}_k$ is a filtering function \cite{Bonanno:2000ep}, $\mathcal{R}_k(p^2) \propto k^2 R^{(0)}(z)$, with $z \equiv p^2/k^2$ and $R^{(0)}(z) \equiv \frac{z}{e^z - 1}$.
Inserting $\Gamma_k$ into (\ref{watt}) and projecting the flow onto the subspace spanned by the Einstein-Hilbert truncation, one infers a system of coupled differential equations for the dimensionless Newton's constant $g(k)\equiv k^{2}G(k)$, and the dimensionless cosmological constant
$\lambda(k)\equiv \bar{\lambda}(k)/k^2$. Since the cosmological constant plays no role within the scope of the present investigation we may approximate $\lambda \approx 0$. Thus the evolution of $g(t)$ is governed by the equation
\begin{align} \label{betag}
\frac{\mathrm{d}g(t)}{\mathrm{d}t} \equiv \beta(g(t)) = \left[ 2 + \frac{B_1 g(t)}{1- B_2 g(t)} \right] \: g(t),
\end{align}
where the constants $B_1$ and $B_2$ are given by \cite{Bonanno:2000ep}
\begin{align}
B_1 \equiv B_1(0) &= -\frac{1}{3\pi} \bigg( 24 \Phi^2_2(0) - \Phi_{1}^{1} (0) \bigg),
\\
B_2 \equiv B_2(0) &= \frac{1}{6\pi} \bigg( 18 \tilde{\Phi}^2_2(0) -5 \tilde{\Phi}_{1}^{1} (0) \bigg).
\end{align}
$\Phi^{p}_{n} (w)$ and $\tilde{\Phi}^{p}_{n} (w)$ are two auxiliary functions defined by
\begin{align}
\Phi^{p}_{n} (w) &\equiv \frac{1}{\Gamma(n)}
\int_{0}^{\infty} \mathrm{d}z \ z^{n-1}
\frac{ R^{(0)}(z)  - z {R^{(0)}}'(z) }{ (z + R^{(0)}(z) + w)^{p} },
\\
\tilde{\Phi}^{p}_{n} (w) &\equiv \frac{1}{\Gamma(n)} \int_{0}^{\infty}
\mathrm{d}z \ z^{n-1}
\frac{ R^{(0)}(z) }{ (z + R^{(0)} (z) + w)^{p} }.
\end{align}
Integrating \eqref{betag} to get an explicit form for $g(k)$, and using the above relation $g(k)=k^2G(k)/\hbar$ (here we restored $\hbar$), one obtains the dimensionfull Newton's coupling
\be \label{basic}
G(k) = \frac{G_0}{1 + \tilde{\omega} G_0 k^2/\hbar}.
\ee
As this equation shows, deviations from the classical solution are important at high energy/momentum scales. The classical space-time is recovered in the opposite regime, when $k\to 0$.
To connect $k$ with the physical radial coordinate $r$, in \cite{Bonanno:2000ep} it has been used the identification
\be \label{kpaper}
k(r) \equiv \hbar \left(\frac{ r + \gamma G_0 M }{r^3}\right)^{1/2} .
\ee
The renormalization scale $k(r)$ is a modified proper distance, more precisely $k(r)\sim \hbar/d(r)$: this function interpolates smoothly the behaviour of the proper distance close to $r=0$ and at infinity. The final result of the modification of the Newtonian gravitational constant is expressed by Eq. (\ref{Geffr}).
%


%
%
%

\begin{thebibliography}{99}

%
%
%
%
%
%
%
%
%
%
%
%
%
%
%
%
%
%
%
%
%
%
%
%
\bibitem{QG1}
T.~Jacobson,
\newblock {\em Phys. Rev. Lett.}, \textbf{75} 1260 (1995).

\bibitem{QG2}
A.~Connes,
\newblock {\em Commun. Math. Phys.}, \textbf{182} 155 (1996).

\bibitem{QG3}
M.~Reuter,
\newblock {\em Phys. Rev.}, D \textbf{57} 971 (1998).

\bibitem{QG4}
C.~Rovelli,
\newblock {\em Living Rev. Rel.}, \textbf{1} 1 (1998).


\bibitem{QG6}
A.~Ashtekar,
\newblock {\em New J. Phys.}, \textbf{7} 198 (2005).


\bibitem{QG8}
P.~Horava,
\newblock {\em Phys. Rev.}, D \textbf{79} 084008 (2009).

\bibitem{QG9}
E.P.~Verlinde,
\newblock {\em JHEP}, \textbf{04} 029 (2011).

\bibitem{Koch:2016uso}
B.~Koch, I.A.~Reyes, \~A.~Rincon,
\newblock {\em Class. Quant. Grav.}, \textbf{33} (22) 225010 (2016).


\bibitem{Rincon:2017goj}
\~A.Rincon, E.Contreras, P.Bargueno, B.Koch, G.Panotopoulos, A.Hernandez-Arboleda,
{\em Eur. Phys. J.}, C \textbf{77} (7) 494 (2017).


\bibitem{Contreras:2017eza}
E.~Contreras, \~A.~Rincon, B.~Koch, P.~Bargueno,
\newblock {\em Int. J. Mod. Phys.}, D \textbf{27} (03) 1850032 (2017).

\bibitem{Rincon:2018sgd}
\~A.~Rincon, G.~Panotopoulos,
\newblock {\em Phys. Rev.}, D \textbf{97} (2) 024027 (2018).

\bibitem{Contreras:2018dhs}
E.~Contreras, \~A.~Rincon, B.~Koch, P.~Bargueno,
\newblock {\em Eur. Phys. J.}, C \textbf{78} (3) 246 (2018).

\bibitem{Rincon:2018lyd}
\~A.~Rincon, B.~Koch,
\newblock {\em Eur. Phys. J.}, C \textbf{78} (12) 1022 (2018).

\bibitem{Rincon:2018dsq}
\~A.~Rincon, E.~Contreras, P.~Bargueno, B.~Koch, G.~Panotopoulos,
\newblock {\em Eur. Phys. J.}, C \textbf{78} (8) 641 (2018).


\bibitem{Rincon:2019cix}
\~A.~Rincon, E.~Contreras, P.~Bargueno, B.~Koch,
\newblock {\em Eur. Phys. J. Plus}, \textbf{134} (11) 557 (2019).



\bibitem{Contreras:2018gpl}
E.~Contreras, P.~Bargueno,
\newblock {\em Mod. Phys. Lett.}, A \textbf{33} (32) 1850184 (2018).

\bibitem{Rincon:2019zxk}
\~A.~Rincon, J.R.~Villanueva,
\newblock Class. Quant. Grav. \textbf{37} (2020), 175003.


\bibitem{Fathi:2019jid}
M.~Fathi, \~A.~Rincon, J.R.~Villanueva,
\newblock Class. Quant. Grav. \textbf{37} (2020), 075004.


\bibitem{Canales:2018tbn}
F.~Canales, B.~Koch, C.~Laporte, \~A.~Rincon.
\newblock JCAP \textbf{01} (2020), 021.

\bibitem{Contreras:2018swc}
E.~Contreras, P.~Bargueno,
\newblock {\em Int. J. Mod. Phys.}, D \textbf{27} (09) 1850101 (2018).

\bibitem{Bonanno:2000ep}
A.~Bonanno, M.~Reuter,
\newblock {\em Phys. Rev.}, D \textbf{62} 043008 (2000).

\bibitem{Bonanno:2001xi}
A.~Bonanno, M.~Reuter,
\newblock {\em Phys. Rev.}, D \textbf{65} 043508 (2002).

\bibitem{Reuter:2003ca}
M.~Reuter, H.~Weyer,
\newblock {\em Phys. Rev.}, D \textbf{69} 104022 (2004).

\bibitem{Platania:2020lqb}
A.Platania,
\newblock {\em Front. in Phys.}, \textbf{8} 188 (2020).


\bibitem{Bonanno:2002zb}
A.Bonanno, M.~Reuter,
\newblock {\em Int. J. Mod. Phys. D}, \textbf{13} 107 (2004).

\bibitem{Bonanno:2001hi}
A.~Bonanno, M.~Reuter,
\newblock {\em Phys. Lett. B}, \textbf{527} 9 (2002).

\bibitem{Liu:2018hno}
L.H.~Liu, T.~Prokopec, A.~Starobinsky,
\newblock {\em Phys. Rev. D}, \textbf{98} (4) 043505 (2018).

\bibitem{Hindmarsh:2012rc}
M.~Hindmarsh, I.D.~Saltas,
\newblock {\em Phys. Rev. D}, \textbf{86} 064029 (2012).

\bibitem{Platania:2019kyx}
A.~Platania,
\newblock {\em Eur. Phys. J. C}, \textbf{79} (6) 470 (2019).

\bibitem{Moti:2018uvl}
R.~Moti, A.~Shojai,
\newblock {\em Int. J. Mod. Phys. A}, \textbf{35} 2050016 (2020).

\bibitem{Koch:2013owa}
B.~Koch, F.~Saueressig,
\newblock {\em Class. Quant. Grav.}, \textbf{31} 015006 (2014).

\bibitem{Bonanno:2016dyv}
A.~Bonanno, B.~Koch, A.~Platania,
\newblock {\em Class. Quant. Grav.}, \textbf{34} (9) 095012 (2017).

%

\bibitem{chinosII}
Dao-Jun Liu, Bin Yang, Yong-Jia Zhai, Xin-Zhou Li,
\newblock {\em Class. Quant. Grav.}, \textbf{29} 145009 (2012).

\bibitem{Li:2013kkb}
Jin Li, Yuanhong Zhong,
\newblock {\em Int. J. Theor. Phys.}, \textbf{52} 1583 (2013).

\bibitem{bardeen}
J.M.~Bardeen,
\textit{Proceedings of of International Conference GR5} (Tbilisi, USSR, 1968) p. 174.

\bibitem{hayward}
S.A.~Hayward,
Phys. Rev. Lett. \textbf{96} (2006), 031103.

\bibitem{frolov}
V.P.~Frolov,
Phys. Rev. D \textbf{94} (2016) no.10, 104056.

\bibitem{Heisenberg}
W.~Heisenberg, Zeitschrift f\"{u}r Physik, {\bf 43}, 172 (1927).
%
\bibitem{GUPearly}

H.S.~Snyder, Phys.Rev. {\bf 71}, 38 (1947);\\
C.N.~Yang, Phys. Rev. {\bf 72}, 874 (1947);\\
C.A.~Mead, Phys. Rev. {\bf 135}, B 849 (1964);\\
F.~Karolyhazy, Nuovo Cim. A {\bf 42}, 390 (1966).
%
\bibitem{VenezGrossMende}
D.~Amati, M.~Ciafaloni, G.~Veneziano, Phys. Lett. B {\bf 197}, 81 (1987);\\
D.J.~Gross, P.F.~Mende, Phys. Lett. B {\bf 197}, 129 (1987);\\
D.~Amati, M.~Ciafaloni, G.~Veneziano, Phys. Lett. B {\bf 216}, 41 (1989);\\
K~.Konishi, G.~Paffuti, P.~Provero, Phys. Lett. B {\bf 234}, 276 (1990).
%
\bibitem{MM}
M.~Maggiore, Phys. Lett. B {\bf 304}, 65 (1993).
%
\bibitem{kempf}
A.~Kempf, G.~Mangano, R.B.~Mann, Phys. Rev. D {\bf 52}, 1108 (1995).
%
\bibitem{FS}
F.~Scardigli, Phys. Lett. B {\bf 452}, 39 (1999).
%

\bibitem{Adler2}
R.J.~Adler, D.I.~Santiago, Mod.~Phys.~Lett. {\bf A14}, 1371 (1999).
%
%
\bibitem{CLS} S. Capozziello, G. Lambiase, G. Scarpetta, Int. J. Theor. Phys. {\bf 39}, 15 (2000).
%
\bibitem{SC}
F.~Scardigli, R.~Casadio, Class.~Quantum Grav. {\bf 20}, 3915 (2003).
%
%
%
%
\bibitem{FS9506} F.~Scardigli, Nuovo Cim. B {\bf 110}, 1029 (1995).
%
\bibitem{ACSantiago}
R.J.~Adler, P.~Chen, D.I.~Santiago, Gen. Rel. Grav. {\bf
33},  2101 (2001).
%
%
\bibitem{CDM03}
M.~Cavaglia, S.~Das and R.~Maartens, Class. Quant. Grav. {\bf 20}, L205 (2003).
%
\bibitem{Susskind}
L.~Susskind, J.~Lindesay, {\em An Introduction to Black Holes,
Information, and the String Theory Revolution} (World Scientific,
Singapore, 2005). See chapter 10.
%
\bibitem{nouicer}
K.~Nouicer, Class. Quant. Grav. {\bf 24}, 5917 (2007).
%
\bibitem{Glimpses}
F.~Scardigli,
Symmetry \textbf{12} (2020) no.9, 1519.
%
%
\bibitem{dvali} G.~Dvali, C.~Gomez,
Fortsch. Phys. \textbf{61}, 742 (2013);\\
R.~Casadio, A.~Giugno, A.~Orlandi, Phys. Rev. D {\bf 91}, 124069 (2015);\\
R.~Casadio, A.~Giugno, A.~Giusti, Phys. Lett. B {\bf 763}, 337 (2016);\\
L.~Buoninfante, G.~G.~Luciano and L.~Petruzziello,
Eur. Phys. J. C \textbf{79} (2019), 663;\\
A.~Giusti,
Int. J. Geom. Meth. Mod. Phys. \textbf{16} (2019), 1930001.
%
%
\bibitem{GUP}
G.~Luciano, L.~Petruzziello,
Eur. Phys. J. C \textbf{79} (2019), 283;\\
L.~Petruzziello, F.~Wagner,
Phys. Rev. D \textbf{103} (2021), 104061;\\
M.~J.~Lake, M.~Miller, R.~F.~Ganardi, Z.~Liu, S.~D.~Liang, T.~Paterek,
Class. Quant. Grav. \textbf{36} (2019), 155012;\\
M.~Bishop, J.~Lee, D.~Singleton,
Phys. Lett. B \textbf{802} (2020), 135209;\\
A. Alonso-Serrano, M.P. Dabrowski, H. Gohar, Phys. Rev. D {\bf 97}, 044029 (2018);\\
F.~Scardigli, M.~Blasone, G.~Luciano, R.~Casadio,
Eur. Phys. J. C \textbf{78} (2018), 728;\\
R.~Casadio, F.~Scardigli,
Phys. Lett. B \textbf{807} (2020), 135558;\\
%
S.~Das, M.~Fridman, G.~Lambiase, E.~Vagenas,
Phys. Lett. B \textbf{824} (2022), 136841;\\
%
R.~Casadio and F.~Scardigli,
Eur. Phys. J. C \textbf{74} (2014) no.1, 2685;\\
%
S.Hassanabadi, J.Kriz, W.S.Chung, B.C.L\"utf\"uo\u{g}lu, E.Maghsoodi, H.Hassanabadi,
Eur.Phys.J.Plus \textbf{136} (2021) no.9, 918;\\
%
I.~Pikovski, C.~Brukner, et al., Nat. Phys. {\bf 8}, 393 (2012).
%
%
%
\bibitem{Hawking}
S.~W.~Hawking,
Commun. Math. Phys. \textbf{43} (1975), 199.
%
\bibitem{SC2} F.~Scardigli, R.~Casadio, Eur. Phys. J. C {\bf 75} 425 (2015).
\bibitem{SLV} F.~Scardigli, G.~Lambiase, E.C.~Vagenas, Phys. Lett. B {\bf 767}, 242 (2017).
\bibitem{LambSME} G. Lambiase, F. Scardigli, Phys.Rev. D {\bf 97}, 075003 (2018).
\bibitem{SLVConf} F.~Scardigli, G.~Lambiase, E.C.~Vagenas, J. Phys. Conf. Ser. \textbf{880}, 012044 (2017).
\bibitem{tomoyo} T. Kanazawa, G. Lambiase, G. Vilasi, A. Yoshioka,
Eur. Phys. J. C {\bf 79}, 95 (2019).
%
%
\bibitem{Duff}  M.J.~Duff,
  Phys.\ Rev.\ D {\bf 9}, 1837 (1974).
\bibitem{donoghue} J.F. Donoghue, Phys. Rev. Lett. {\bf 72}, 2996 (1994);
Phys. Rev. D {\bf 50}, 3874 (1994).	
\bibitem{donoASP}
J.F.~Donoghue,
Front. in Phys. \textbf{8}, 56 (2020).
\bibitem{kiril}
H.W.~Hamber, S.~Liu, Phys.Lett.B {\bf 357}, 51 (1995);\\
I.B.~Khriplovich, G.G.~Kirilin, J.Exp.Theor.Phys. {\bf 95}, 981 (2002);\\
I.B.~Khriplovich, G.G.~Kirilin, J.Exp.Theor.Phys. {\bf 98}, 1063 (2004);\\
N.E.J.~Bjerrum-Bohr, J.F.~Donoghue, B.R.~Holstein, Phys.Rev.D {\bf 68}, 084005  (2003); Erratum-ibid.D {\bf 71}, 069904 (2005).
\bibitem{dono2}
N.E.J.~Bjerrum-Bohr, J.F.~Donoghue, B.R.~Holstein, Phys.Rev.D {\bf 67}, 084033 (2003); Erratum-ibid.D {\bf 71},  069903 (2005);\\
A.Akhundov, A.Shiekh, EJTP \textbf{5}, No. 17, 1 (2008);\\
C.~Kiefer, J. Phys. Conf. Ser. \textbf{442}, 012025 (2013);\\
N.E.J.~Bjerrum-Bohr, J.F.~Donoghue, B.R.~Holstein, L.~Plant\'e, P.~Vanhove,
Phys. Rev. Lett. \textbf{114}, 061301 (2015);\\
J.F.~Donoghue, B.R.~Holstein, J. Phys. G \textbf{42} 10, 103102 (2015).
%
%
%
%
\bibitem{SLnew}
F.~Scardigli, G.~Lambiase, {\em in preparation}.
%
%
\bibitem{JKS}
P.~Jizba, H.~Kleinert, F.~Scardigli, Phys. Rev. D {\bf 81}, 084030 (2010).
%
%
\bibitem{ong}
Y.C.~Ong,
JCAP \textbf{09} (2018), 015.
%
%
\bibitem{jizba2022} P. Jizba, G. Lambiase, G.Luciano, L. Petruzziello [arXiv:2201.07919]
%
%
\bibitem{Bushev}
P.A.~Bushev, J.~Bourhill, M.~Goryachev, N.~Kukharchyk, E.~Ivanov, S.~Galliou, M.E.~Tobar, S.~Danilishin,
Phys. Rev. D \textbf{100} (2019) no.6, 066020.
%
%
%
%

\bibitem{martin} F.~Marin, M.~Cerdonio et al., Nat. Phys. {\bf 9}, 71 (2013).
\bibitem{ghosh} S. Ghosh, Class. Quantum Gravity \textbf{31}, 025025 (2014).
\bibitem{Bonaldi} M. Bawaj, C. Biancofiore, F. Marin et al., Nat. Commun. {\bf 6}, 7503 (2015).

\bibitem{vagenas}
A.F.~Ali, S.~Das, E.C.~Vagenas, Phys.~Rev.~D {\bf 84}, 044013 (2011);\\
S.~Das, E.C.~Vagenas Can. J. Phys. {\bf 84}, 233 (2009).
%

\bibitem{Tkachuk} C.~Quesne, V.M.~Tkachuk, Phys. Rev. A {\bf 81}, 012106 (2010).
\bibitem{gao} D. Gao, M. Zhan, Phys. Rev. A {\bf 94}, 013607 (2016).
\bibitem{ron} J. Steinhauer, Nature Physics {\bf 10}, 864 (2014);\\
R. Cowen, Nature News, 12 October 2014, {\it Hawking radiation mimicked in the lab},  (doi:10.1038/nature.2014.16131).




\end{thebibliography}
\end{document}